# Supporting Material for

# Target Atmospheric CO$_2$: Where Should Humanity Aim?

James Hansen[1],[*],[a,b], Makiko Sato[a,b], Pushker Kharecha[a,b], David Beerling[c], Robert Berner[d], Valerie Masson-Delmotte[e], Mark Pagani[d], Maureen Raymo[f], Dana L. Royer[g] and James C. Zachos[h]

[a]*NASA/Goddard Institute for Space Studies, New York, NY 10025, USA*
[b]*Columbia University Earth Institute, New York, NY 10027, USA*
[c]*Dept. Animal and Plant Sciences, University of Sheffield, Sheffield S10 2TN, UK*
[d]*Dept. Geology and Geophysics, Yale University, New Haven, CT 06520-8109, USA*
[e]*Lab. Des Sciences du Climat et l'Environnement/Institut Pierre Simon Laplace, CEA-CNRS-Universite de Versailles Saint-Quentin en Yvelines, CE Saclay, 91191, Gif-sur-Yvette, France*
[f]*Dept. Earth Sciences, Boston University, Boston, MA 02215, USA*
[g]*Dept. Earth and Environmental Sciences, Wesleyan University, Middletown, CT 06459-0139, USA*
[h]*Earth & Planetary Sciences Dept., University of California, Santa Cruz, Santa Cruz, CA 95064, USA*

**Abstract:** Paleoclimate data show that climate sensitivity is ~3°C for doubled CO$_2$, including only fast feedback processes. Equilibrium sensitivity, including slower surface albedo feedbacks, is ~6°C for doubled CO$_2$ for the range of climate states between glacial conditions and ice-free Antarctica. Decreasing CO$_2$ was the main cause of a cooling trend that began 50 million years ago, large scale glaciation occurring when CO$_2$ fell to 450 ± 100 ppm, a level that will be exceeded within decades, barring prompt policy changes. If humanity wishes to preserve a planet similar to that on which civilization developed and to which life on Earth is adapted, paleoclimate evidence and ongoing climate change suggest that CO$_2$ will need to be reduced from its current 385 ppm to at most 350 ppm. The largest uncertainty in the target arises from possible changes of non-CO$_2$ forcings. An initial 350 ppm CO$_2$ target may be achievable by phasing out coal use except where CO$_2$ is captured and adopting agricultural and forestry practices that sequester carbon. If the present overshoot of this target CO$_2$ is not brief, there is a possibility of seeding irreversible catastrophic effects.

---

[*]Address correspondence to this author at NASA/Goddard Institute for Space Studies, New York, NY 10025, USA; E-mail: jhansen@giss.nasa.gov

# Appendix: Supporting Material

## 1. Ice age climate forcings

Fig. (**S1**) shows the climate forcings during the depth of the last ice age, 20 ky BP, relative to the Holocene [14]. The largest contribution to the uncertainty in the calculated 3.5 W/m$^2$ forcing due to surface changes (ice sheet area, vegetation distribution, shoreline movements) is due to uncertainty in the ice sheet sizes [14, S1]. Formulae for the GHG forcings [20] yield 2.25 W/m$^2$ for $CO_2$ (185 ppm → 275 ppm), 0.43 W/m$^2$ for $CH_4$ (350 → 675 ppb) and 0.32 W/m$^2$ for $N_2O$ (200 → 270 ppb). The $CH_4$ forcing includes a factor 1.4 to account for indirect effects of $CH_4$ on tropospheric ozone and stratospheric water vapor [12].

The climate sensitivity inferred from the ice age climate change (~¾°C per W/m$^2$) includes only fast feedbacks, such as water vapor, clouds, aerosols (including dust) and sea ice. Ice sheet size and greenhouse gas amounts are specified boundary conditions in this derivation of the fast-feedback climate sensitivity.

It is permissible, alternatively, to specify aerosol changes as part of the forcing and thus derive a climate sensitivity that excludes the effect of aerosol feedbacks. That approach was used in the initial empirical derivation of climate sensitivity from Pleistocene climate change [14]. The difficulty with that approach is that, unlike long-lived GHGs, aerosols are distributed heterogeneously, so it is difficult to specify aerosol changes accurately. Also the forcing is a sensitive function of aerosol single scatter albedo and the vertical distribution of aerosols in the atmosphere, which are not measured. Furthermore, the aerosol indirect effect on clouds also depends upon all of these poorly known aerosol properties.

One recent study [S2] specified an arbitrary glacial-interglacial aerosol forcing slightly larger than the GHG glacial-interglacial forcing. As a result, because temperature, GHGs, and aerosol amount, overall, are positively correlated in glacial-interglacial changes, this study inferred a climate sensitivity of only ~2°C for doubled $CO_2$. This study used the correlation of aerosol and temperature in the Vostok ice core at two specific times to infer an aerosol forcing for a given aerosol amount. The conclusions of the study are immediately falsified by considering the full Vostok aerosol record (Fig. **2** of [17]), which reveals numerous large aerosol fluctuations without any corresponding temperature change. In contrast, the role of GHGs in climate change is confirmed when this same check is made for GHGs (Fig. **2**), and the fast-feedback climate sensitivity of 3°C for doubled $CO_2$ is confirmed (Fig. **1**).

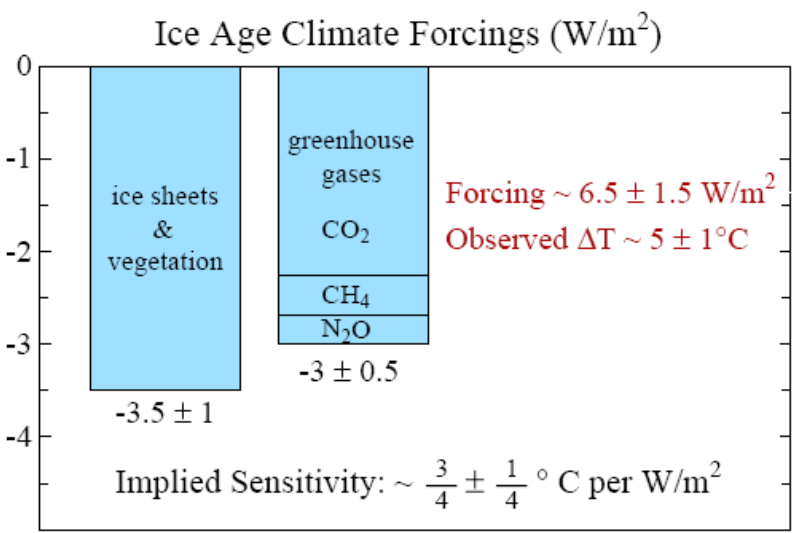

**Fig. (S1).** Climate forcings during ice age 20 ky BP, relative to the present (pre-industrial) interglacial period.

All the problems associated with imprecise knowledge of aerosol properties become moot if, as is appropriate, aerosols are included in the fast feedback category. Indeed, soil dust, sea salt, dimethylsulfide, and other aerosols are expected to vary (in regional, inhomogeneous ways) as climate changes. Unlike long-lived GHGs, global aerosol amounts cannot be inferred from ice cores. But the effect of aerosol changes is fully included in observed global temperature change. The climate sensitivity that we derive in Fig. **(S1)** includes the aerosol effect accurately, because both the climate forcings and the global climate response are known. The indirect effect of aerosol change on clouds is, of course, also included precisely.

## 2. Climate forcings and climate feedbacks.

The Earth's temperature at equilibrium is such that the planet radiates to space (as heat, i.e., infrared radiation) the same amount of energy that it absorbs from the sun, which is ~240 W/m$^2$. A blackbody temperature of ~255°K yields a heat flux of 240 W/m$^2$. Indeed, 255°K is the temperature in the mid-troposphere, the mean level of infrared emission to space.

A climate forcing is a perturbation to the planet's energy balance, which causes the Earth's temperature to change as needed to restore energy balance. Doubling atmospheric $CO_2$ causes a planetary energy imbalance of ~4 W/m$^2$, with more energy coming in than going out. Earth's temperature would need to increase by $\Delta T_O$ = 1.2-1.3°C to restore planetary energy balance, if the temperature change were uniform throughout the atmosphere and if nothing else changed.

Actual equilibrium temperature change in response to any forcing is altered by feedbacks that can amplify or diminish the response, thus the mean surface temperature change is [14]

$$\begin{aligned}\Delta T_{eq} &= f \, \Delta T_O \\ &= \Delta T_O + \Delta T_{feedbacks} \\ &= \Delta T_O + \Delta T_1 + \Delta T_2 + \ldots,\end{aligned}$$

where f is the net feedback factor and the $\Delta T_i$ are increments due to specific feedbacks.

The role of feedback processes is clarified by defining the gain, g,

$$\begin{aligned}g &= \Delta T_{feedbacks}/\Delta T_{eq} \\ &= (\Delta T_1 + \Delta T_2 + \ldots)/\Delta T_{eq} \\ &= g_1 + g_2 + \ldots\end{aligned}$$

$g_i$ is positive for an amplifying feedback and negative for a feedback that diminishes the response. The additive nature of the $g_i$, unlike $f_i$, is a useful characteristic of the gain. Evidently

$$f = 1/(1 - g)$$

The value of g (or f) depends upon the climate state, especially the planetary temperature. For example, as the planet becomes so warm that land ice disappears, the land ice albedo feedback diminishes, i.e. $g_{land\,ice\,albedo} \rightarrow 0$.

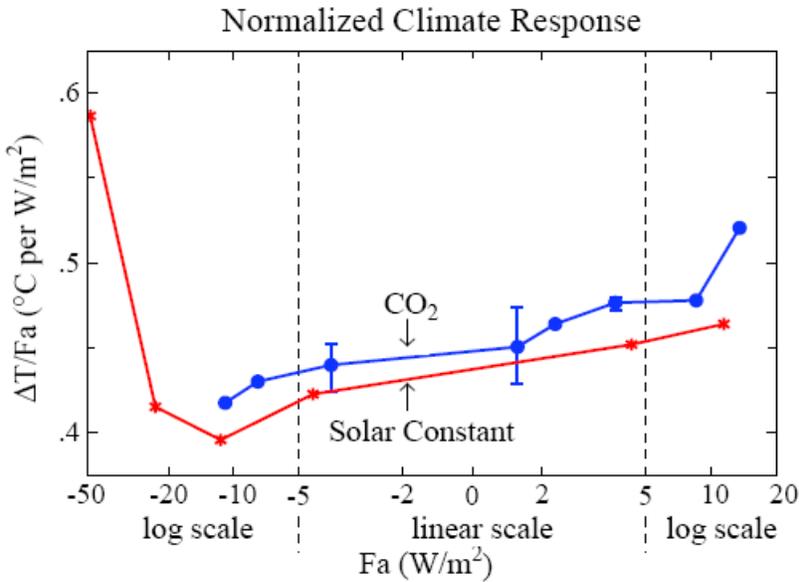

**Fig. (S2).** Global surface air temperature change [12] after 100 years in simulations with the Goddard Institute for Space Studies modelE [S3, 5] as a function of climate forcing for changes of solar irradiance and atmospheric $CO_2$. Fa is the standard adjusted climate forcing [12]. Results are extracted from Fig. **(25a)** of [12]. Curves terminate because the climate model 'bombs' at the next increment of forcing due to failure of one or more of the parameterizations of processes in the model as extreme conditions are approached.

'Fast feedbacks', such as water vapor, clouds and sea ice, are the mechanisms usually included in the 'Charney' [13] climate sensitivity. Climate models yield a Charney (fast feedback) sensitivity of about 3°C for doubled $CO_2$ [2, 12], a conclusion that is confirmed and tightened by empirical evidence from the Pleistocene (Section 2.1). This sensitivity implies

$$g_{\text{fast feedbacks}} \sim 0.5\text{-}0.6.$$

This fast feedback gain and climate sensitivity apply to the present climate and climate states with global temperatures that are not too different than at present.

If g approaches unity, f → ∞, implying a runaway climate instability. The possibility of such instability is anticipated for either a very warm climate (runaway greenhouse effect [S4]) or a very cold climate (snowball Earth [S5]). We can investigate how large a climate forcing is needed to cause g → 1 using a global climate model that includes the fast feedback processes, because both of these instabilities are a result of the temperature dependence of 'fast feedbacks' (the water vapor and ice/snow albedo feedbacks, respectively).

Fig. **(S2)** suggests that climate forcings ~10-25 W/m$^2$ are needed to approach either runaway snowball-Earth conditions or the runaway greenhouse effect. More precise quantification requires longer simulations and improved parameterizations of physical processes as extreme climates are approached. The processes should include slow feedbacks that can either amplify or diminish the climate change.

Earth has experienced snowball conditions [S5], or at least a 'slushball' state [S6] with ice reaching sea level in the tropics, on at least two occasions, the most recent ~640 My BP, aided by reduced solar irradiance [43] and favorable continental locations. The mechanism that allowed Earth to escape the snowball state was probably reduced weathering in a glaciated world, which allowed $CO_2$ to accumulate in the atmosphere [S5]. Venus, but not Earth, has experienced the runaway greenhouse effect, a state from which there is no escape.

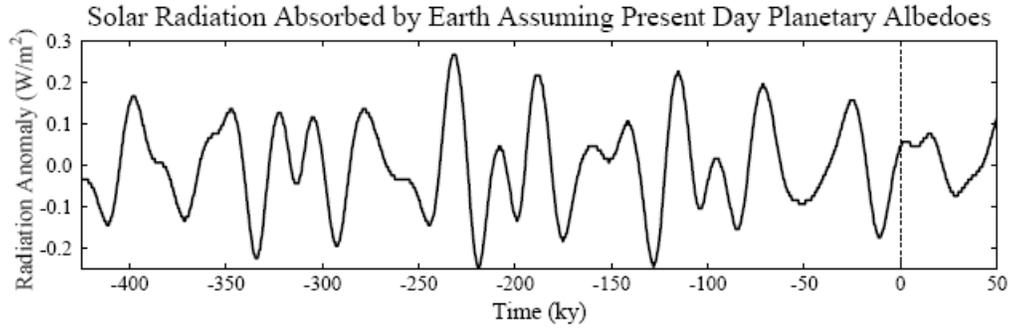

**Fig. (S3).** Annual-mean global-mean perturbation of the amount of solar radiation absorbed by the Earth, calculated by assuming present-day seasonal and geographical distribution of albedo.

## 3. Pleistocene forcings and feedbacks

Fig. **(S3)** shows the perturbation of solar radiation absorbed by the Earth due to changes in Earth orbital elements, i.e., the tilt of the Earth's spin axis relative to the orbital plane, the eccentricity of the Earth's orbit, and the time of year at which the Earth is closest to the sun (precession of equinoxes). This perturbation is calculated using fixed (present day) seasonal and geographical distribution of planetary albedo.

The global-mean annual-mean orbital (Milankovitch) forcing is very weak, at most a few tenths of 1 W/m$^2$. Our procedure in calculating the forcing, keeping ice sheet properties (size and albedo) fixed, is appropriate for 'instantaneous' and 'adjusted' radiative forcings [12].

Further, successive, definitions of the orbital 'forcing', e.g., allowing some regional response to the seasonal insolation perturbations, may be useful for the purpose of understanding glacial-interglacial climate change. For example, it may be informative to calculate the 'forcing' due to insolation-induced changes of ice-sheet albedo, because increased insolation can 'age' (increase snow crystal size and thus darken) an ice surface and also spur the date of first snow-melt [7]. However, one merit of the standard forcing definition is the insight that glacial-interglacial climate swings are almost entirely due to feedbacks.

Indeed, the gain during the Pleistocene is close to unity. Climate models and empirical evaluation from the climate change between the last ice age (Section 2.1 above) yield $g_{\text{fast feedbacks}}$ ~0.5-0.6 (the gain corresponding to fast feedback climate sensitivity 3°C for doubled $CO_2$). GHGs and surface albedo contribute about equally to glacial-interglacial 'forcings' and temperature change, with each having gain ~0.2 [14]. Thus

$$g = g_{\text{fast feedbacks}} + g_{\text{surface albedo}} + g_{\text{GHG}}$$
$$= \sim 0.5\text{-}0.6 + \sim 0.2 + \sim 0.2 \ .$$

Thus climate gain in the Pleistocene was greater than or of the order of 0.9. It is no wonder that late Cenozoic climate fluctuated so greatly (Fig. 3b). When substantial ice is present on the planet, g is close to unity, climate is sensitive, and large climate swings occur in response to small orbital forcings. Indeed, with g near unity any forcing or climate noise can cause large climate change, consistent with the conclusion that much of climate variability is not due to orbital forcings [S7]. In the early Cenozoic there was little ice, $g_{\text{surface albedo}}$ was small, and thus climate oscillations due to insolation perturbations were smaller.

It may be useful to divide inferences from Pleistocene climate change into two categories: (1) well-defined conclusions about the nature of the climate change, (2) less certain suggestions about the nature and causes of the climate change. The merit of identifying well-defined conclusions is that they help us predict likely consequences of human-made climate forcings. Less certain aspects of Pleistocene climate change mainly concern the small forcings that instigated climate swings. The small forcings are of great interest to paleoclimatologists, but they need not prevent extraction of practical implications from Pleistocene climate change.

Two fundamental characteristics of Pleistocene climate change are clear. First, there is the high gain, at least of the order of 0.9, i.e., the high sensitivity to a climate forcing, when the planet is in the range of climates that existed during the Pleistocene. Second, we have a good knowledge of the amplifying feedbacks that produce this high gain. Fast feedbacks, including water vapor, clouds, aerosols, sea ice and snow, contribute at least half of this gain. The remainder of the amplification is provided almost entirely by two factors: surface albedo (mainly ice sheets) and GHGs (mainly $CO_2$).

Details beyond these basic conclusions are less certain. The large glacial-interglacial surface albedo and GHG changes should lag global temperature, because they are feedbacks on global temperature on the global spatial scale and millennial time scale. The lag of GHGs after temperature change is several hundred years [Fig. 6 of 6], perhaps determined by the ocean overturning time. Ice sheet changes may lag temperature by a few millennia [24], but it has been argued that there is no discernible lag between insolation forcing and the maximum rate of change of ice sheet volume [7].

A complication arises from the fact that some instigating factors (forcing mechanisms) for Pleistocene climate change also involve surface albedo and GHG changes. Regional anomalies of seasonal insolation are as much as many tens of $W/m^2$. The global forcing is small (Fig. S3) because the local anomalies are nearly balanced by anomalies of the opposite sign in either the opposite hemisphere or the opposite season. However, one can readily imagine climate change mechanisms that operate in such a way that cancellation does not occur.

For example, it has been argued [7] that a positive insolation anomaly in late spring is most effective for causing ice sheet disintegration because early 'albedo flip', as the ice becomes wet, yields maximum extension of the melt season. It is unlikely that the strong effect of albedo flip on absorbed solar energy could be offset by a negative insolation anomaly at other times of year.

A second example is non-cancellation of hemispheric insolation anomalies. A hemispheric asymmetry occurs when Earth is cold enough that ice sheets extend to Northern Hemisphere middle latitudes, due to absence of similar Southern Hemisphere land. It has been argued [7] that this hemispheric asymmetry is the reason that the orbital periodicities associated with precession of the equinoxes and orbit eccentricity became substantial about 1 million years ago.

Insolation anomalies also may directly affect GHG amounts, as well as surface albedo. One can readily imagine ways in which insolation anomalies affect methane release from wetlands or carbon uptake through biological processes.

Surface albedo and GHG changes that result immediately from insolation anomalies can be defined as climate forcings, as indirect forcings due to insolation anomalies. The question is then: what fractions of known paleo albedo and GHG changes are immediate indirect forcings due to insolation anomalies and what fractions are feedbacks due to global temperature change?

It is our presumption that most of the Pleistocene GHG changes are a slow feedback in response to climate change. This interpretation is supported by the lag of several hundred years between temperature change and greenhouse gas amount [Fig. 6 of 6]. The conclusion that most of the ice area and surface albedo change is also a feedback in response to global temperature change is supported by the fact that the large climate swings are global (Section 5 of Appendix).

Note that our inferred climate sensitivity is not dependent on detailed workings of Pleistocene climate fluctuations. The fast feedback sensitivity of 3°C for doubled $CO_2$, derived by comparing glacial and interglacial states, is independent of the cause and dynamics of glacial/interglacial transitions.

Climate sensitivity including surface albedo feedback (~6°C for doubled $CO_2$) is the average sensitivity for the climate range from 35 My ago to the present and is independent of the glacial-interglacial 'wiggles' in Fig. 3.  Note that climate and albedo changes occurred mainly at points with 'ready' [63] feedbacks: at Antarctic glaciation and (in the past three million years) with expansion of Northern Hemisphere glaciation, which are thus times of high climate sensitivity.

The entire ice albedo feedback from snowball-Earth to ice-free planet (or vice versa) can be viewed as a response to changing global temperature, with wiggles introduced by Milankovitch (orbital) forcings.  The average $g_{surface\ albedo}$ for the range from today's climate through Antarctic deglaciation is close to $g_{surface\ albedo}$ ~ 0.2, almost as large as in the Pleistocene.  Beyond Antarctic deglaciation (i.e., for an ice-free planet) $g_{surface\ albedo} \rightarrow 0$, except for vegetation effects.

For the sake of specificity, let us estimate the effect of slow feedbacks on climate sensitivity.  If we round $\Delta T_O$ to 1.2°C for doubled $CO_2$ and the fast feedback gain to $g_{fast\ feedbacks}$ = 0.6, then for fast feedbacks alone f = 2.5 and the equilibrium warming is $\Delta T_{eq}$ = 3°C.  Inclusion of $g_{surface\ albedo}$ = 0.2 makes f = 5 and $\Delta T_{eq}$ = 6°C, which is the sensitivity if the GHG amount is specified from observations or from a carbon cycle model.

The feedback factor f can approach infinity, i.e., the climate can become unstable.  However, instabilities are limited except at the snowball Earth and runaway greenhouse extremes.  Some feedbacks have a finite supply, e.g., as when Antarctica becomes fully deglaciated.  Also climate change can cause positive feedbacks to decrease or negative feedbacks to come into play.

For example, Fig. S2 suggests that a cooling climate from the present state first reduces the fast feedback gain.  This and reduced weathering with glaciation may be reasons that most ice ages did not reach closer to the iceball state.  Also there may be limitations on the ranges of GHG ($CO_2$, $CH_4$, $N_2O$) feedbacks.  Empirical values $g_{GHG}$ ~ 0.2 and $g_{surface\ albedo}$ ~ 0.2 were derived as averages relevant to the range of climates that existed in the past several hundred thousand years, and they may not be valid outside that range.

On the other hand, if the forcing becomes large enough, global instabilities are possible.  Earth did become cold enough in the past for the snowball-Earth instability.  Although the runaway greenhouse effect has not occurred on Earth, solar irradiance is now at its highest level so far, and Fig. S2 suggests that the required forcing for runaway may be only 10-20 W/m$^2$.  If all conventional and unconventional fossil fuels were burned, with the $CO_2$ emitted to the atmosphere, it is possible that a runaway greenhouse effect could occur, with incineration of life and creation of a permanent Venus-like hothouse Earth.  It would take time for the ice sheets to melt, but the melt rate may accelerate as ice sheet disintegration proceeds.

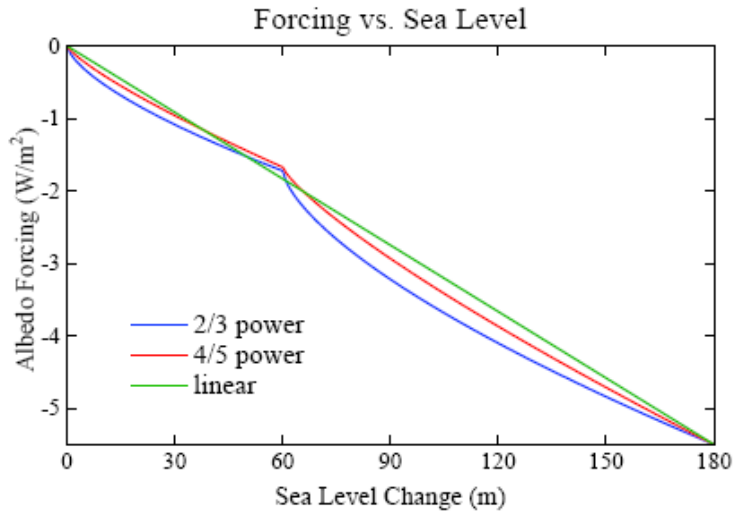

**Fig. (S4).** Surface albedo climate forcing as a function of sea level for three approximations of the ice sheet area as a function of sea level change, from an ice free planet to the last glacial maximum.  For sea level between 0 and 60 m only Antarctica contributes to the albedo change.  At the last glacial maximum Antarctica contains 75 m of sea level and the Northern Hemisphere contains 105 m.

## 4. Ice sheet albedo

In the present paper we take the surface area covered by an ice sheet to be proportional to the 4/5 power of the volume of the ice sheet, based on ice sheet modeling of one of us (VM-D). We extend the formulation all the way to zero ice on the planet, with separate terms for each hemisphere. At 20 ky ago, when the ice sheets were at or near their maximum size in the Cenozoic era, the forcing by the Northern Hemisphere ice sheet was -3.5 W/m$^2$ and the forcing by the Southern Hemisphere ice sheet was -2 W/m$^2$, relative to the ice-free planet [14]. It is assumed that the first 60 m of sea level fall went entirely into growth of the Southern Hemisphere ice sheet. The water from further sea level fall is divided proportionately between hemispheres such that when sea level fall reaches -180 m there is 75 m in the ice sheet of the Southern Hemisphere and 105 m in the Northern Hemisphere.

The climate forcing due to sea level changes in the two hemispheres, $SL_S$ and $SL_N$, is

$$F_{Albedo} (W/m^2) = -2 (SL_S/75 \text{ m})^{4/5} - 3.5 (SL_N/105 \text{ m})^{4/5}, \qquad (S1)$$

where the climate forcings due to fully glaciated Antarctica (-2 W/m$^2$) and Northern Hemisphere glaciation during the last glacial maximum (-3.5 W/m$^2$) were derived from global climate model simulations [14].

Fig. **(S4)** compares results from the present approach with results from the same approach using exponent 2/3 rather than 4/5, and with a simple linear relationship between the total forcing and sea level change. Use of exponent 4/5 brings the results close to the linear case, suggesting that the simple linear relationship is a reasonably good approximation. The similarity of Fig. **(1c)** in our present paper and Fig. **(2c)** in [7] indicates that change of exponent from 2/3 to 4/5 did not have a large effect.

## 5. Global nature of major climate changes

Climate changes often begin in a specific hemisphere, but the large climate changes are invariably global, in part because of the global GHG feedback. Even without the GHG feedback, forcings that are located predominately in one hemisphere, such as ice sheet changes or human-made aerosols, still evoke a global response [12], albeit with the response being larger in the hemisphere of the forcing. Both the atmosphere and ocean transmit climate response between hemispheres. The deep ocean can carry a temperature change between hemispheres with little loss, but because of the ocean's thermal inertia there can be a hemispheric lag of up to a millennium (see Ocean Response Time, below).

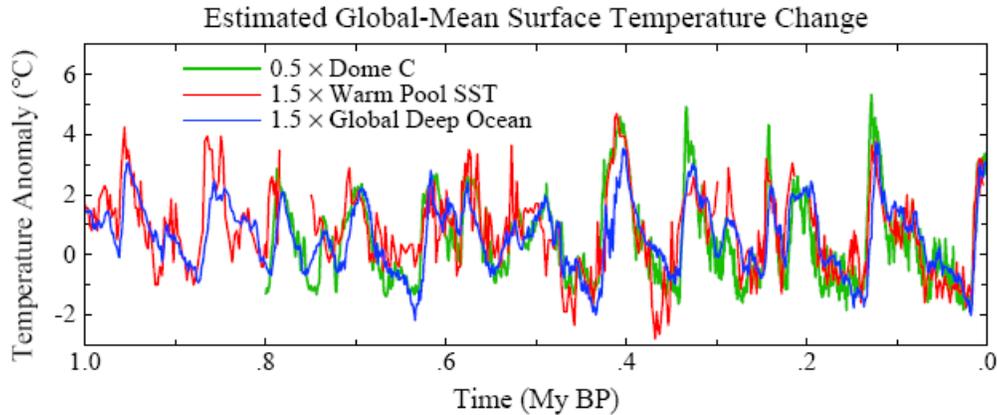

**Fig. (S5).** Estimated global temperature change based on measurements at a single point or, in the case of the deep ocean, a near-global stack of ocean drilling sites: Antarctica Dome C (S8), Warm Pool (S9), deep ocean (26).

Fig. **(S5)** compares temperature change in Antarctica [S8], the tropical sea surface [S9], and the global deep ocean [26]. Temperature records are multiplied by factors that convert the temperature record to an estimate of global temperature change. Based on paleoclimate records, polar temperature change is typically twice the global average temperature change, and tropical temperature change is about two-thirds of the global mean change. This polar amplification of the temperature change is an expected consequence of feedbacks [14], especially the snow-ice albedo feedback. The empirical result that deep ocean temperature changes are only about two-thirds as large as global temperature change is obtained from data for the Pleistocene epoch, when deep ocean temperature change is limited by its approach to the freezing point.

## 6. Holocene climate forcings

The GHG zero-point for the paleo portion of Fig. **(2)** is the mean for 10-8 ky BP, a time that should precede any significant anthropogenic effect on GHG amount. It has been suggested that the increase of $CO_2$ that began 8000 years ago is due to deforestation and the increase of $CH_4$ that began 6000 years ago is caused by rice agriculture [62]. This suggestion has proven to be controversial, but regardless of whether late Holocene $CO_2$ and $CH_4$ changes are human-made, the GHG forcing is anomalous in that period relative to global temperature change estimated from ocean and ice cores. As discussed elsewhere [7], the late Holocene is the only time in the ice core record in which there is a clear deviation of temperature from that expected due to GHG and surface albedo forcings.

The GHG forcing increase in the second half of the Holocene is ~3/4 W/m$^2$. Such a large forcing, by itself, would create a planetary energy imbalance that could not be sustained for millennia without causing a large global temperature increase, the expected global warming being about 1°C. Actual global temperature change in this period was small, perhaps a slight cooling. Fig. **(S6)** shows estimates of global temperature change obtained by dividing polar temperature change by two or multiplying tropical and deep ocean temperatures by 1.5. Clearly the Earth has not been warming rapidly in the latter half of the Holocene. Thus a substantial (negative) forcing must have been operating along with the positive GHG forcing.

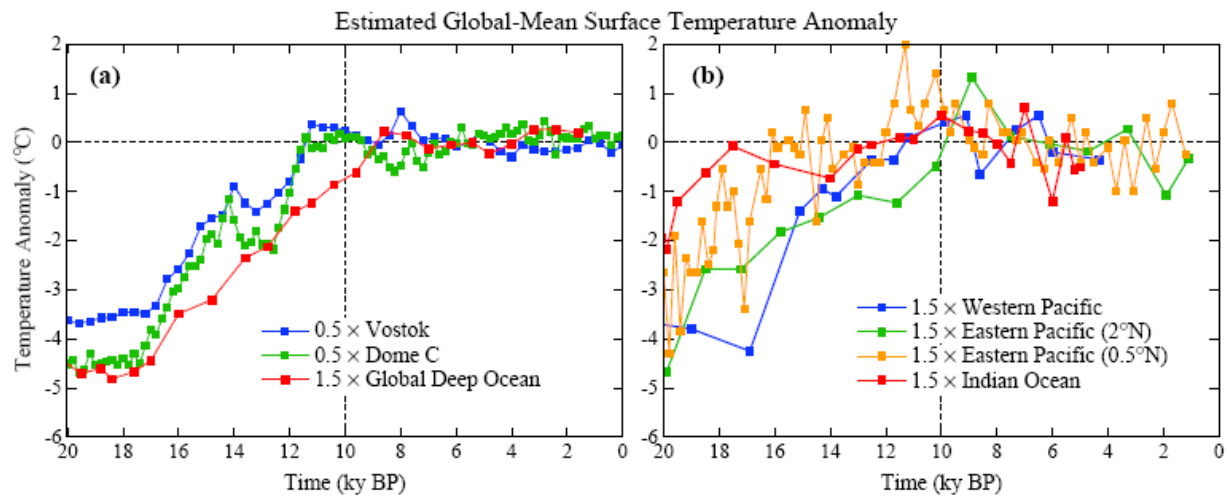

**Fig. (S6).** Estimates of global temperature change inferred from Antarctic ice cores [18, S8] and ocean sediment cores [S9-S13], as in Fig. **(S5)** but for a period allowing Holocene temperature to be apparent.

Deforestation causes a negative climate forcing [12], but an order of magnitude too small to balance GHG positive forcing. A much larger negative forcing is expected from human-made aerosols. Aerosol forcing is non-linear, especially the indirect effect on clouds, with aerosols added to a pristine atmosphere being more effective than those added to the current highly polluted atmosphere. Given estimates of a negative forcing of 1-2 W/m$^2$ for today's anthropogenic aerosols [2, 5, 12], a negative aerosol forcing at least of the order of 0.5 W/m$^2$ in 1850 is expected. We conclude that aerosols probably were the predominant negative forcing that opposed the rapid increase of positive GHG forcing in the late Holocene.

**7. Ocean response time**

Fig. **(S7)** shows the climate response function, defined as the fraction of equilibrium global warming that is obtained as a function of time. This response function was obtained [7] from a 3000-year simulation after instant doubling of atmospheric $CO_2$, using GISS modelE [S3, 12] coupled to the Russell ocean model [S14]. Note that although 40% of the equilibrium solution is obtained within several years, only 60% is achieved after a century, and nearly full response requires a millennium. The long response time is caused by slow uptake of heat by the deep ocean, which occurs primarily in the Southern Ocean.

This delay of the surface temperature response to a forcing, caused by ocean thermal inertia, is a strong (quadratic) function of climate sensitivity and it depends on the rate of mixing of water into the deep ocean [31]. The ocean model used for Fig. **(S7)** may mix somewhat too rapidly in the waters around Antarctica, as judged by transient tracers [S14], reducing the simulated surface response on the century time scale. However, this uncertainty does not qualitatively alter the shape of the response function (Fig. **S7**).

When the climate model used to produce Fig. **(S7)** is driven by observed changes of GHGs and other forcings it yields good agreement with observed global temperature and ocean heat storage [5]. The model has climate sensitivity ~3°C for doubled $CO_2$, in good agreement with the fast-feedback sensitivity inferred from paleoclimate data.

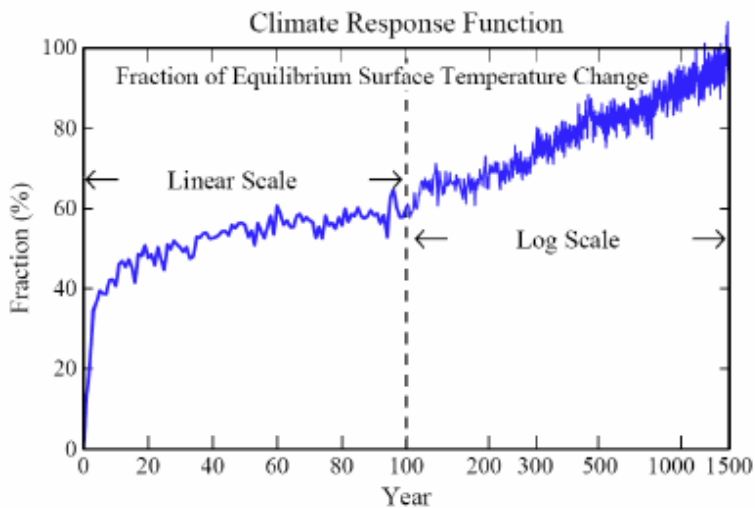

**Fig. (S7).** Fraction of equilibrium surface temperature response versus time in the GISS climate model [7, 12, S3] with the Russell [S14] ocean. The forcing was doubled atmospheric $CO_2$. The ice sheets, vegetation distribution and other long-lived GHGs were fixed.

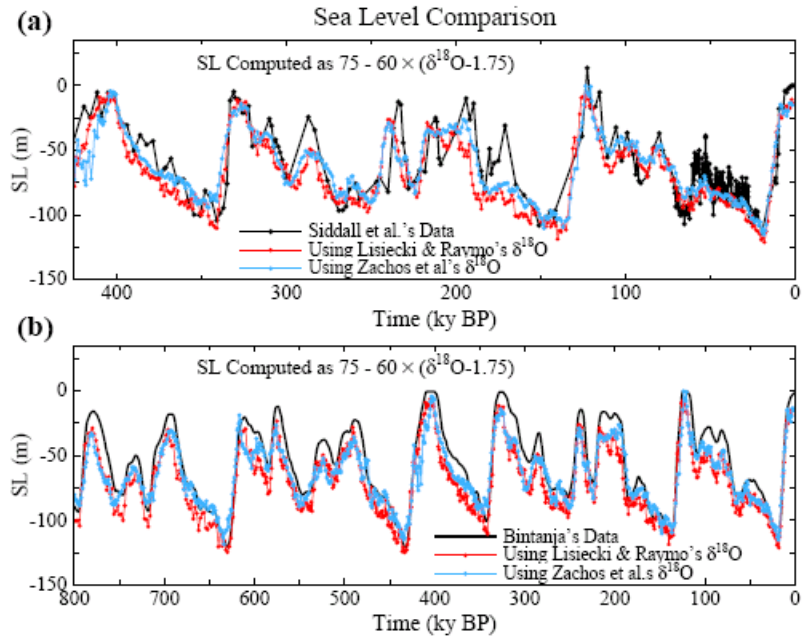

**Fig. (S8).**
(a) Comparison of Siddall et al. [19] sea level record with sea level computed from $\delta^{18}O$ via Eq. S2 using two alternative global benthic stacks [26, S15].
(b) Comparison of Bintanja et al. [S16] sea level reconstruction with the same global benthic stacks as in (a).

## 8. Separation of $\delta^{18}O$ into ice volume and temperature

$\delta^{18}O$ of benthic (deep ocean dwelling) foraminifera is affected by both deep ocean temperature and continental ice volume. Between 34 My and the last ice age (20 ky) the change of $\delta^{18}O$ was ~ 3, with $T_{do}$ change ~ 6°C (from +5 to -1°C) and ice volume change ~ 180 msl (meters of sea level). Based on the rate of change of $\delta^{18}O$ with deep ocean temperature in the prior period without land ice, ~ 1.5 of $\delta^{18}O$ is associated with the $T_{do}$ change of ~ 6°C, and we assign the remaining $\delta^{18}O$ change to ice volume linearly at the rate 60 msl per mil $\delta^{18}O$ change (thus 180 msl for $\delta^{18}O$ between 1.75 and 4.75).

Thus we assume that ice sheets were absent when $\delta^{18}O < 1.75$ with sea level 75 msl higher than today. Sea level at smaller values of $\delta^{18}O$ is given by

$$SL\ (m) = 75 - 60 \times (\delta^{18}O - 1.75). \qquad (S2)$$

Fig. **(S8)** shows that the division of $\delta^{18}O$ equally into sea level change and deep ocean temperature captures well the magnitude of the major glacial to interglacial changes.

## 9. Continental drift and atmospheric $CO_2$

At the beginning of the Cenozoic era 65 My ago the continents were already close to their present latitudes, so the effect of continental location on surface albedo had little direct effect on the planet's energy balance (Fig. **S9**). However, continental drift has a major effect on the balance, or imbalance, of outgassing and uptake of $CO_2$ by the solid Earth and thus a major effect on atmospheric composition and climate. We refer to the carbon in the air, ocean, soil and biosphere as the combined surface reservoir of carbon, and carbon in ocean sediments and the rest of the crust as the carbon in the 'solid' Earth. Sloshing of $CO_2$ among the surface reservoirs, as we have shown, is a primary mechanism for glacial-interglacial climate fluctuations. On longer time scales the total amount of carbon in the surface reservoirs can change as a result of any imbalance between outgassing and uptake by the solid Earth.

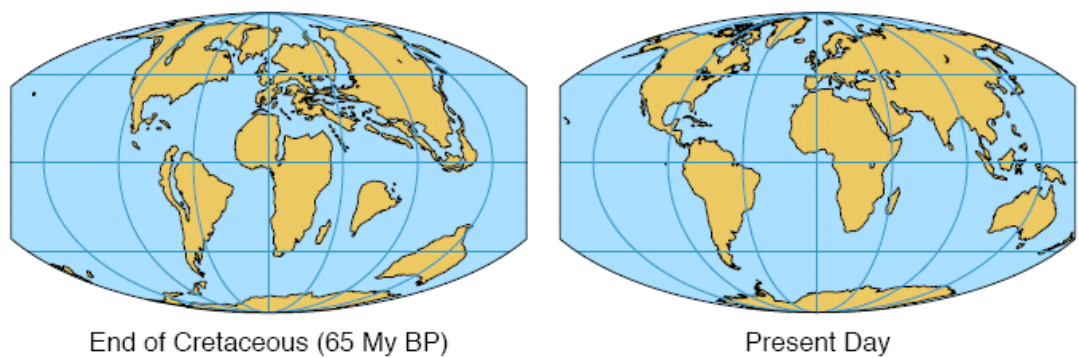

**Fig. S9.** Continental locations at the beginning and end of the Cenozoic era [S17].

Outgassing, which occurs mainly in regions of volcanic activity, depends upon the rate at which carbon-rich oceanic crust is subducted beneath moving continental plates [30, 47]. Drawdown of $CO_2$ from the surface reservoir occurs with weathering of rocks exposed by uplift, with the weathering products carried by rivers to the ocean and eventually deposited as carbonates on the ocean floor [30] and by burial of organic matter. Both outgassing and drawdown of $CO_2$ are affected by changes in plate tectonics, which thus can alter the amount of carbon in the surface reservoir. The magnitude of the changes of carbon in the surface reservoir, and thus in the atmosphere, is constrained by a negative weathering feedback on the time scale of hundreds of thousands of years [30, 52], but plate tectonics can evoke changes of the surface carbon reservoir by altering the rates of outgassing and weathering.

At the beginning of the Cenozoic the African plate was already in collision with Eurasia, pushing up the Alps. India was still south of the equator, but moving north rapidly through a region with fresh carbonate deposits. It is likely that subduction of carbon rich crust of the Tethys Ocean, long a depocenter for sediments, caused an increase of atmospheric $CO_2$ and the early Cenozoic warming that peaked ~50 My ago. The period of rapid subduction terminated with the collision of India with Eurasia, whereupon uplift of the Himalayas and the Tibetan Plateau increased weathering rates and drawdown of atmospheric $CO_2$ [51].

Since 50 My ago the world's major rivers have emptied into the Indian and Atlantic Oceans, but there is little subduction of oceanic crust of these regions that are accumulating sediments [47]. Thus the collision of India with Asia was effective in both reducing a large source of outgassing of $CO_2$ as well as exposing rock for weathering and drawdown of atmospheric $CO_2$. The rate of $CO_2$ drawdown decreases as the $CO_2$ amount declines because of negative feedbacks, including the effects of temperature and plant growth rate on weathering [30].

## 10. Proxy $CO_2$ data

There are inconsistencies among the several proxy measures of atmospheric $CO_2$, including differences between results of investigators using nominally the same reconstruction method. We briefly describe strengths and weaknesses of the four paleo-$CO_2$ reconstruction methods included in the IPCC report [2], which are shown in Fig. **(S10)** and discussed in detail elsewhere [S18]. The inconsistencies among the different proxies constrain their utility for rigorously evaluating our $CO_2$ predictions. We also include a comparison of our calculated $CO_2$ history with results from a version of the Berner [30] geochemical carbon cycle model, as well as a comparison with an emerging $CO_2$ proxy based on carbon-isotope analyses of nonvascular plant (bryophyte) fossils [S19].

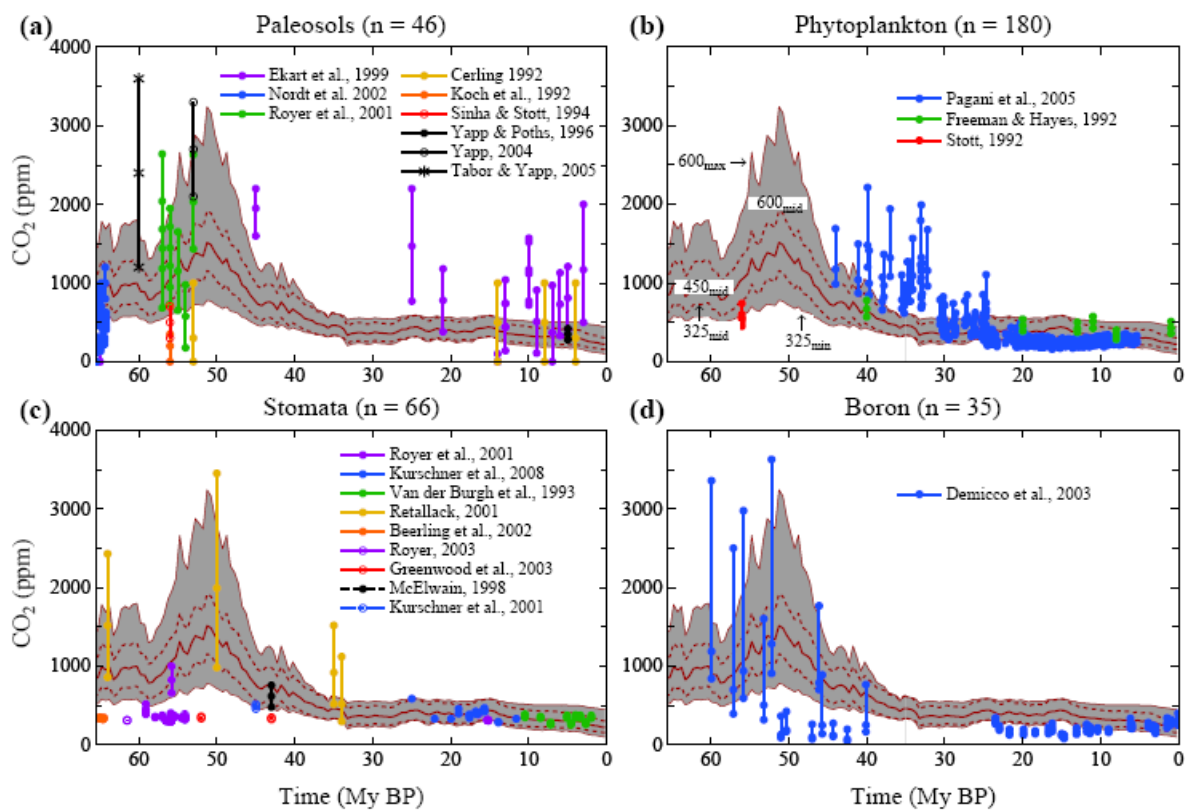

**Fig. (S10).** Comparison of proxy $CO_2$ measurements with $CO_2$ predictions based on deep-ocean temperature, the latter inferred from benthic $\delta^{18}O$. The shaded range of model results is intended mainly to guide the eye in comparing different proxies. The dark central line is for the standard case with $CO_2 = 450$ ppm at 35 My ago, and the dashed lines are the standard cases for $CO_2 = 325$ and 600 ppm at 35 My ago. The extremes of the shaded area correspond to the maximum range including a 50% uncertainty in the relation of $\Delta T_s$ and $\Delta T_{do}$. Our assumption that $CO_2$ provides 75% of the GHG throughout the Cenozoic adds additional uncertainty to the predicted $CO_2$ amount. References for data sources in the legends are provided by Royer [55], except Kurshner *et al.* [S20].

The paleosol method is based on the $\delta^{13}C$ of pedogenic carbonate nodules, whose formation can be represented by a two end-member mixing model between atmospheric $CO_2$ and soil-derived carbon [S21]. Variables that need to be constrained or assumed include an estimation of nodule depth from the surface of the original soil, the respiration rate of the ecosystem that inhabits the soil, the porosity/diffusivity of the original soil, and the isotopic composition of the vegetation contribution of respired $CO_2$. The uncertainties in $CO_2$ estimates with this proxy are substantial at high $CO_2$ (±500-1000 ppm when $CO_2 > 1000$ ppm) and somewhat less in the lower $CO_2$ range (±400-500 ppm when $CO_2 < 1000$ ppm).

The stomatal method is based on the genetically-controlled relationship [S22] between the proportion of leaf surface cells that are stomata and atmospheric $CO_2$ concentrations [S23]. The error terms with this method are comparatively small at low $CO_2$ (< ±50 ppm), but the method rapidly loses sensitivity at high $CO_2$ (> 500-1000 ppm). Because stomatal-$CO_2$ relationships are often species-specific, only extant taxa with long fossil records can be used [S24]. Also, because the fundamental response of stomata is to the partial pressure of $CO_2$ [S25], constraints on paleoelevation are required.

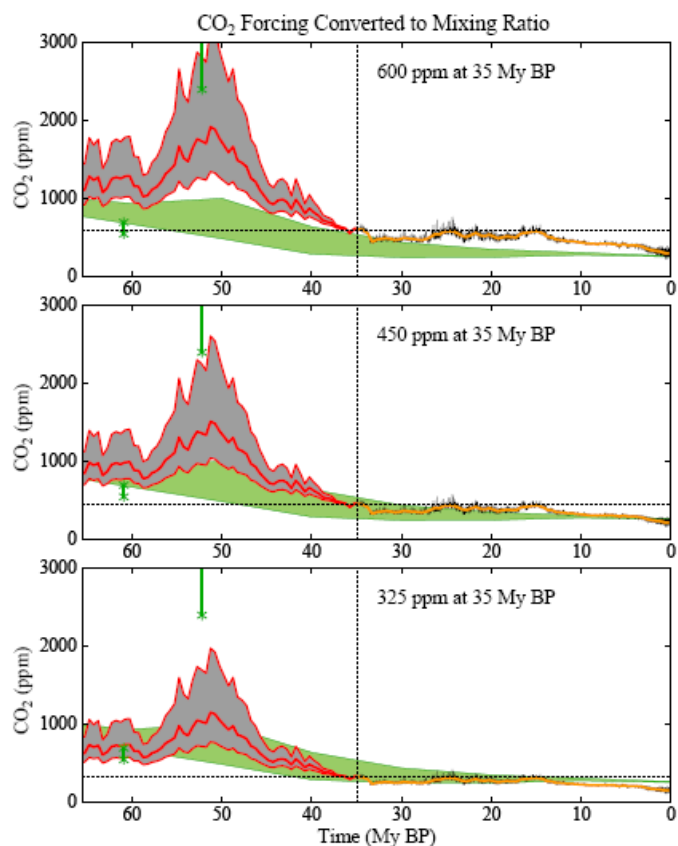

**Fig. (S11).** Simulated $CO_2$ in the Cenozoic for three choices of $CO_2$ amount at 35 My, as in Fig. **(5)**, compared with the $CO_2$ history in a geochemical model [30], specifically the model version described by Fletcher *et al.* [S19]. The green vertical bars are a proxy $CO_2$ measure [S19] obtained from fossils of non-vascular plants (bryophytes) that is not included among the proxies shown in Fig. **(S10)**.

The phytoplankton method is based on the Rayleigh distillation process of fractionating stable carbon isotopes during photosynthesis [S26]. In a high $CO_2$ environment, for example, there is a higher diffusion rate of $CO_2$ through phytoplankton cell membranes, leading to a larger available intercellular pool of $CO_{2[aq]}$ and more depleted $\delta^{13}C$ values in photosynthate. Cellular growth rate and cell size also impact the fractionation of carbon isotopes in phytoplankton and thus fossil studies must take these factors into account [S27]. This approach to reconstructing $CO_2$ assumes that the diffusional transport of $CO_2$ into the cell dominates, and that any portion of carbon actively transported into the cell remains constant with time. Error terms are typically small at low $CO_2$ ($< \pm 50$ ppm) and increase substantially under higher $CO_2$ concentrations [S27].

The boron-isotope approach is based on the pH-dependency of the $\delta^{11}B$ of marine carbonate [S28]. This current method assumes that only borate is incorporated in the carbonate lattice and that the fractionation factor for isotope exchange between boric acid and borate in solution is well-constrained. Additional factors that must be taken into account include test dissolution and size, species-specific physiological effects on carbonate $\delta^{11}B$, and ocean alkalinity [S29-S31]. As with the stomatal and phytoplankton methods, error terms are comparatively small at low $CO_2$ ($< \pm 50$ ppm) and the method loses sensitivity at higher $CO_2$ ($> 1000$ ppm). Uncertainty is unconstrained for extinct foraminiferal species.

Fig. **(S10)** illustrates the scatter among proxy data sources, which limits inferences about atmospheric $CO_2$ history. Given the large inconsistency among different data sets in the early Cenozoic, at least some of the data or their interpretations must be flawed. In the range of proxy data shown in Fig. **(5)** we took all data sources as being of equal significance. It seems likely

that the low $CO_2$ values in the early Cenozoic are faulty, but we avoid omission of any data until the matter is clarified, and thus the range of proxy data shown in Fig. **(5)** is based on all data. Reviews of the proxy data [S19, 55] conclude that atmospheric $CO_2$ amount in the early Cenozoic reached values of at least 500-1000 ppm.

Fig. (**S11**) shows that geochemical carbon cycle modeling [30, S19] is reasonably consistent with our calculated long-term trend of atmospheric $CO_2$ for the cases with $CO_2$ at 34 My ago being in the range from about 325 to 450 ppm. The geochemical modeling does not yield a strong maximum of $CO_2$ at 50 My ago, but the temporal resolution of the modeling (10 My) and the absence of high resolution input data for outgassing due to variations in plate motions tends to mitigate against sharp features in the simulated $CO_2$.

Fig. (**S11**) also shows (vertical green bars) an emerging $CO_2$ proxy based on the isotopic composition of fossil liverworts. These non-vascular plants, lacking stomatal pores, have a carbon isotopic fractionation that is strongly $CO_2$ dependent, reflecting the balance between $CO_2$ uptake by photosynthesis and inward $CO_2$ diffusion [S19].

## 11. Climate sensitivity comparisons

Other empirical or semi-empirical derivations of climate sensitivity from paleoclimate data (Table S1) are in reasonable accord with our results, when account is taken of differences in definitions of sensitivity and the periods considered.

Royer et al. [56] use a carbon cycle model, including temperature dependence of weathering rates, to find a best-fit doubled $CO_2$ sensitivity of 2.8°C based on comparison with Phanerozoic $CO_2$ proxy amounts. Best-fit in their comparison of model and proxy $CO_2$ data is dominated by the times of large $CO_2$ in the Phanerozoic, when ice sheets would be absent, not by the times of small $CO_2$ in the late Cenozoic. Their inferred sensitivity is consistent with our inference of ~3°C for doubled $CO_2$ at times of little or no ice on the planet.

Higgins and Schrag [57] infer climate sensitivity of ~4°C for doubled $CO_2$ from the temperature change during the Paleocene-Eocene Thermal Maximum (PETM) ~55 My ago (Fig. **3**), based on the magnitude of the carbon isotope excursion at that time. Their climate sensitivity for an ice-free planet is consistent with ours within uncertainty ranges. Furthermore, recalling that we assume non-$CO_2$ to provide 25% of the GHG forcing, if one assumes that part of the PETM warming was a direct of effect of methane, then their inferred climate sensitivity is in even closer agreement with ours.

Pagani et al. [58] also use the magnitude of the PETM warming and the associated carbon isotopic excursion to discuss implications for climate sensitivity, providing a graphical relationship to help assess alternative assumptions about the origin and magnitude of carbon release. They conclude that the observed PETM warming of about 5°C implies a high climate sensitivity, but with large uncertainty due to imprecise knowledge of the carbon release.

**Table S1. Climate sensitivity inferred semi-empirically from Cenozoic or Phanerozoic climate change.**

| Reference | Period | Doubled $CO_2$ Sensitivity |
|---|---|---|
| Royer et al. [56] | 0-420 My | ~ 2.8°C |
| Higgins and Schrag [57] | PETM | ~4°C |
| Pagani et al. [58] | PETM | High |

## 12. Greenhouse gas growth rates

Fossil fuel $CO_2$ emissions have been increasing at a rate close to the highest IPCC [S34] scenario (Fig. **S12b**). Increase of $CO_2$ in the air, however, appears to be in the middle of the IPCC scenarios (Fig. **S12c, d**), but as yet the scenarios are too close and interannual variability too large, for assessment. $CO_2$ growth is well above the "alternative scenario", which was defined with the objective of keeping added GHG forcing in the 21$^{st}$ century at about 1.5 W/m$^2$ and 21$^{st}$ century global warming less than 1°C [20].

Non-$CO_2$ greenhouse gases are increasing more slowly than in IPCC scenarios, overall at approximately the rate of the "alternative scenario", based on a review of data through the end of 2007 [69]. There is potential to reduce non-$CO_2$ forcings below the alternative scenario [69].

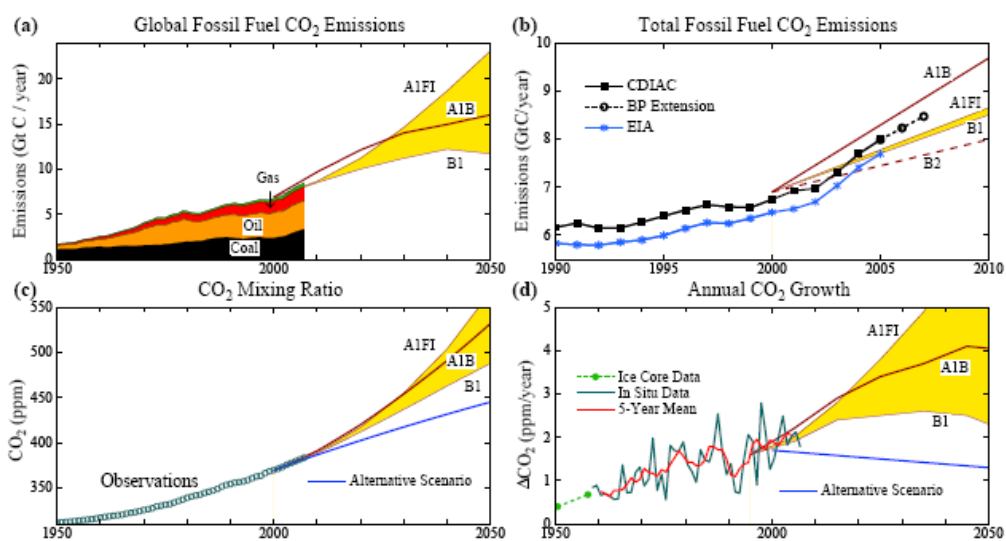

**Fig. (S12). (a)** Fossil fuel $CO_2$ emissions by fuel type [S32, S33], the thin green sliver being gas flaring plus cement production, and IPCC fossil fuel emissions scenarios, **(b)** expansion global emissions to show recent changes more precisely, the EIA values excluding $CO_2$ emissions from cement manufacture, **(c)** observed atmospheric $CO_2$ amount and IPCC and "alternative" scenarios for the future, **(d)** annual atmospheric $CO_2$ growth rates. Data here is an update of data sources defined in [6]. The yellow area is bounded by scenarios that are most extreme in the second half of the 21$^{st}$ century; other scenarios fall outside this range in the early part of the century.

## 13. Fossil fuel and land-use $CO_2$ emissions

Fig. **(S13)** shows estimates of anthropogenic $CO_2$ emissions to the atmosphere. Although fossil emissions through 2006 are known with good accuracy, probably better than 10%, reserves and potential reserve growth are highly uncertain. IPCC [S34] estimates for oil and gas proven reserves are probably a lower limit for future oil and gas emissions, but they are perhaps a feasible goal that could be achieved via a substantial growing carbon price that discourages fossil fuel exploration in extreme environments together with national and international policies that accelerate transition to carbon-free energy sources and limit fossil fuel extraction in extreme environments and on government controlled property.

Coal reserves are highly uncertain, but the reserves are surely enough to take atmospheric $CO_2$ amount far into the region that we assess as being "dangerous". Thus we only consider scenarios in which coal use is phased out as rapidly as possible, except for uses in which the $CO_2$ is captured and stored so that it cannot escape to the atmosphere. Thus the magnitude of coal reserves does not appreciably affect our simulations of future atmospheric $CO_2$ amount.

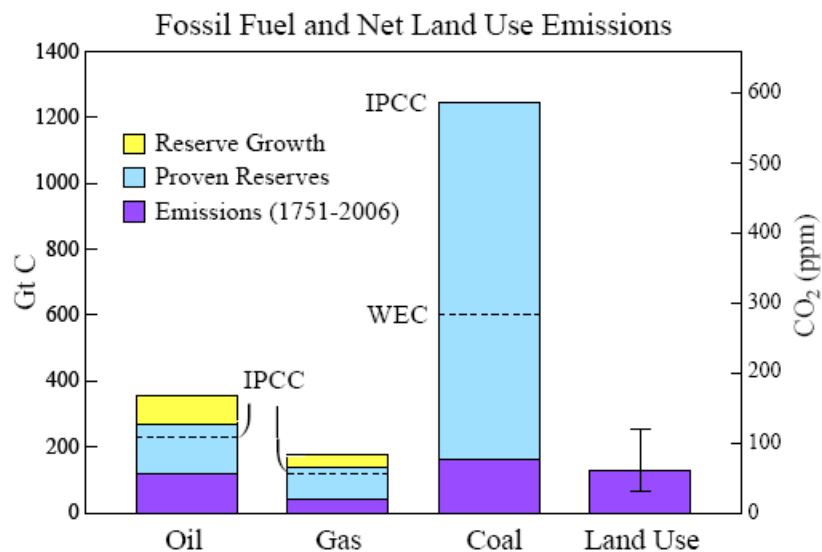

**Fig. (S13).** Fossil fuel and land-use $CO_2$ emissions, and potential fossil fuel emissions. Historical fossil fuel emissions are from the Carbon Dioxide Information Analysis Center [CDIAC, S32] and British Petroleum [BP, S33]. Lower limits on oil and gas reserves are from IPCC [S34] and higher limits are from the United States Energy Information Administration [EIA, 80]. Lower limit for coal reserves is from the World Energy Council [WEC, S35] and upper limit from IPCC [S34]. Land use estimate is from integrated emissions of Houghton/2 (Fig. **S14**) supplemented to include pre-1850 and post-2000 emissions; uncertainty bar is subjective.

Integrated 1850-2008 net land-use emissions based on the full Houghton [83] historical emissions (Fig. **S14**), extended with constant emissions for the past several years, are 79 ppm $CO_2$. Although this could be an overestimate by up to a factor of two (see below), substantial pre-1850 deforestation must be added in. Our subjective estimate of uncertainty in the total land-use $CO_2$ emission is a factor of two.

## 14. The modern carbon cycle.

Atmospheric $CO_2$ amount is affected significantly not only by fossil fuel emissions, but also by agricultural and forestry practices. Quantification of the role of land-use in the uptake and release of $CO_2$ is needed to assess strategies to minimize human-made climate effects.

Fig. **(S15)** shows the $CO_2$ airborne fraction, AF, the annual increase of atmospheric $CO_2$ divided by annual fossil fuel $CO_2$ emissions. AF is a critical metric of the modern carbon cycle, because it is based on the two numbers characterizing the global carbon cycle that are well known. AF averages 56% over the period of accurate data, which began with the $CO_2$ measurements of Keeling in 1957, with no discernable trend. The fact that 44% of fossil fuel emissions seemingly "disappears" immediately provides a hint of optimism with regard to the possibility of stabilizing, or reducing, atmospheric $CO_2$ amount.

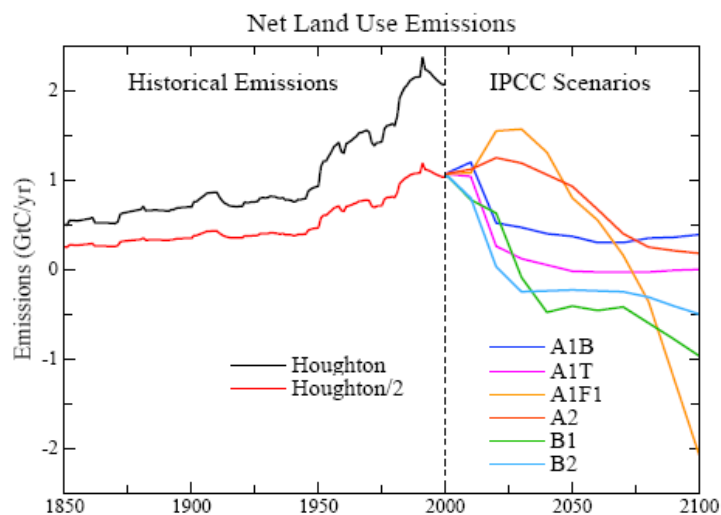

**Fig. (S14).** Left side: estimate by Houghton [83] of historical net land-use $CO_2$ emissions, and a 50 percent reduction of that estimate. Right side: IPCC [2] scenarios for land-use $CO_2$ emissions.

That optimism needs to be tempered, as we will see, by realization of the magnitude of the actions required to halt and reverse $CO_2$ growth. However, it is equally important to realize that assertions that fossil fuel emissions must be reduced close to 100% on an implausibly fast schedule are not necessarily valid.

A second definition of the airborne fraction, AF2, is also useful. AF2 includes the net anthropogenic land-use emission of $CO_2$ in the denominator. This AF2 definition of airborne fraction has become common in recent carbon cycle literature. However, AF2 is not an observed or accurately known quantity; it involves estimates of net land-use $CO_2$ emissions, which vary among investigators by a factor of two or more [2].

Fig. **(S15)** shows an estimate of net land-use $CO_2$ emissions commonly used in carbon cycle studies, labeled "Houghton" [83], as well as "Houghton/2", a 50% reduction of these land-use emissions. An over-estimate of land-use emissions is one possible solution of the long-standing "missing sink" problem that emerges when the full "Houghton" land-use emissions are employed in carbon cycle models [2, S34, 79].

Principal competing solutions of the "missing sink" paradox are (1) land-use $CO_2$ emissions are over-estimated by about a factor of two, or (2) the biosphere is being "fertilized" by anthropogenic emissions, via some combination of increasing atmospheric $CO_2$, nitrogen deposition, and global warming, to a greater degree than included in typical carbon cycle models. Reality may include contributions from both candidate explanations. There is also a possibility that imprecision in the ocean uptake of $CO_2$, or existence of other sinks such as clay formation, could contribute increased $CO_2$ uptake, but these uncertainties are believed to be small.

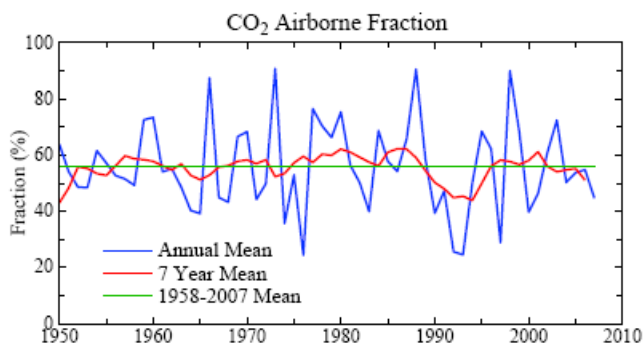

**Fig. (S15).** $CO_2$ airborne fraction, AF, the ratio of annual observed atmospheric $CO_2$ increase to annual fossil fuel $CO_2$ emissions.

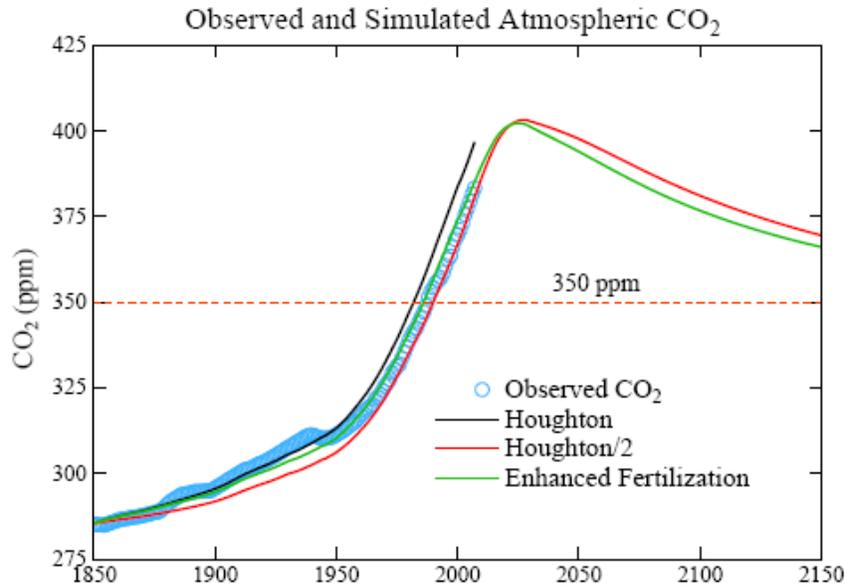

**Fig. (S16).** Computed and observed time evolution of atmospheric $CO_2$. "Enhanced Fertilization" uses the full "Houghton" land use emissions for 1850–2000. "Houghton/2" and "Enhanced Fertilization" simulations are extended to 2100 assuming coal phase-out by 2030 and the IPCC [2] A1T land-use scenario. Observations are from Law Dome ice core data and flask and in-situ measurements [6, S36, http://www.esrl.noaa.gov/gmd/ccgg/trends/].

Fig. **(S16)** shows resulting atmospheric $CO_2$, and Fig. **(S17)** shows AF and AF2, for two extreme assumptions: "Houghton/2" and " Enhanced Fertilization", as computed with a dynamic-sink pulse response function (PRF) representation of the Bern carbon cycle model [78, 79]. Fertilization is implemented via a parameterization [78] that can be adjusted to achieve an improved match between observed and simulated $CO_2$ amount. In the "Houghton/2" simulation

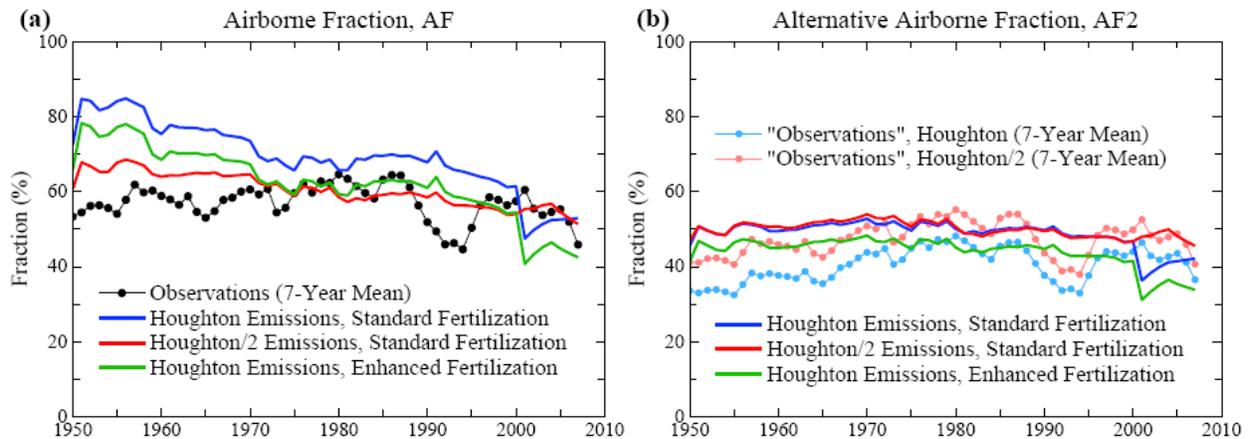

**Fig. (S17).** (a) Observed and simulated airborne fraction (AF), the ratio of annual $CO_2$ increase in the air over annual fossil fuel $CO_2$ emissions, (b) AF2 includes the sum of land use and fossil fuel emissions in the denominator in defining airborne fraction; thus AF2 is not accurately known because of the large uncertainty in land use emissions.

the original value [78] of the fertilization parameter is employed while in the "Enhanced Fertilization" simulation the full Houghton emissions are used with a larger fertilization parameter. Both "Houghton/2" and "Enhanced Fertilization" yield good agreement with the observed $CO_2$ history, but Houghton/2 does a better job of matching the time dependence of observed AF.

It would be possible to match observed $CO_2$ to an arbitrary precision if we allowed the adjustment to "Houghton" land-use to vary with time, but there is little point or need for that. Fig. **(S16)** shows that projections of future $CO_2$ do not differ much even for the extremes of Houghton/2 and Enhanced Fertilization. Thus in Fig. **(6)** we show results for only the case Houghton/2, which is in better agreement with the airborne fraction and also is continuous with IPCC scenarios for land use.

## 15. Implications of Fig. (6): $CO_2$ Emissions and Atmospheric Concentration with Coal Phase-out by 2030.

Fig. **(6)** provides an indication of the magnitude of actions that are needed to return atmospheric $CO_2$ to a level of 350 ppm or lower. Fig. **(6)** allows for the fact that there is disagreement about the magnitude of fossil fuel reserves, and that the magnitude of useable reserves depends upon policies.

A basic assumption underlying Fig. **(6)** is that, within the next several years, there will be a moratorium on construction of coal-fired power plants that do not capture and store $CO_2$, and that $CO_2$ emissions from existing power plants will be phased out by 2030. This coal emissions phase out is the sine qua non for stabilizing and reducing atmospheric $CO_2$. If the sine qua non of coal emissions phase-out is achieved, atmospheric $CO_2$ can be kept to a peak amount ~400-425 ppm, depending upon the magnitude of oil and gas reserves.

Fig. **(6)** illustrates two widely different assumptions about the magnitude of oil and gas reserves (illustrated in Fig. **S13**). The smaller oil and gas reserves, those labeled "IPCC", are realistic if "peak oil" advocates are more-or-less right, i.e., if the world has already exploited about half of readily accessible oil and gas deposits, so that production of oil and gas will begin to decline within the next several years.

There are also "resource optimists" who dispute the "peakists', arguing that there is much more oil (and gas) to be found. It is possible that both the "peakists" and "resource optimists" are right, it being a matter of how hard we work to extract maximum fossil fuel resources. From the standpoint of controlling human-made climate change, it does not matter much which of these parties is closer to the truth.

Fig. **(6)** shows that, if peak $CO_2$ is to be kept close to 400 ppm, the oil and gas reserves actually exploited need to be close to the "IPCC" reserve values. In other words, if we phase out coal emissions we can use remaining oil and gas amounts equal to those which have already been used, and still keep peak $CO_2$ at about 400 ppm. Such a limit is probably necessary if we are to retain the possibility of a drawdown of $CO_2$ beneath the 350 ppm level by methods that are more-or-less "natural". If, on the other hand, reserve growth of the magnitude that EIA estimates (Figs. **6** and **S13**) occurs, and if these reserves are burned with the $CO_2$ emitted to the atmosphere, then the forest and soil sequestration that we discuss would be inadequate to achieve drawdown below the 350 ppm level in less than several centuries.

Even if the greater resources estimated by EIA are potentially available, it does not mean that the world necessarily must follow the course implied by EIA estimates for reserve growth. If a sufficient price is applied to carbon emissions it will discourage extraction of fossil fuels in the most extreme environments. Other actions that would help keep effective reserves close to the IPCC estimates would include prohibition of drilling in environmentally sensitive areas, including the Arctic and Antarctic.

National policies, in most countries, have generally pushed to expand fossil fuel reserves as much as possible. This might partially account for the fact that energy information agencies, such as the EIA in the United States, which are government agencies, tend to forecast strong growth of fossil fuel reserves. On the other hand, state, local, and citizen organizations can influence imposition of limits on fossil fuel extraction, so there is no guarantee that fossil resources will be fully exploited. Once the successors to fossil energy begin to take hold, there may be a shifting away from fossil fuels that leaves some of the resources in the ground. Thus a scenario with oil and gas emissions similar to that for IPCC reserves may be plausible.

Assumptions yielding the Forestry & Soil wedge in Fig. **(6b)** are as follows. It is assumed that current net deforestation will decline linearly to zero between 2010 and 2015. It is assumed that uptake of carbon via reforestation will increase linearly until 2030, by which time reforestation will achieve a maximum potential sequestration rate of 1.6 GtC per year [S37]. Waste-derived biochar application will be phased in linearly over the period 2010-2020, by which time it will reach a maximum uptake rate of 0.16 GtC/yr [85]. Thus after 2030 there will be an annual uptake of $1.6 + 0.16 = 1.76$ GtC per year, based on the two processes described.

Thus Fig. **(6)** shows that the combination of (1) moratorium and phase-out of coal emissions by 2030, (2) policies that effectively keep fossil fuel reserves from significantly exceeding the IPCC reserve estimates, and (3) major programs to achieve carbon sequestration in forests and soil, can together return atmospheric $CO_2$ below the 350 ppm level before the end of the century.

The final wedge in Fig. **(6)** is designed to provide an indication of the degree of actions that would be required to bring atmospheric $CO_2$ back to the level of 350 ppm by a time close to the middle of this century, rather than the end of the century. This case also provides an indication of how difficult it would be to compensate for excessive coal emissions, if the world should fail to achieve a moratorium and phase-out of coal as assumed as our "sine qua non".

Assumptions yielding the Oil-Gas-Biofuels wedge in Fig. **(6b)** are as follows: energy efficiency, conservation, carbon pricing, renewable energies, nuclear power and other carbon-free energy sources, and government standards and regulations will lead to decline of oil and gas emissions at 4% per year beginning when 50% of the estimated resource (oil or gas) has been exploited, rather than the 2% per year baseline decline rate [79]. Also capture of $CO_2$ at gas-power plants (with $CO_2$ capture) will use 50% of remaining gas supplies. Also a linear phase-in of liquid biofuels is assumed between 2015 and 2025 leading to a maximum global bioenergy from "low-input/high-diversity" biofuels of ~23 EJ/yr, inferred from Tilman et al. [87], that is used as a substitute for oil; this is equivalent to ~0.5 GtC/yr, based on energy conversion of 50 EJ/GtC for oil. Finally, from 2025 onward, twice this number (i.e., 1 GtC/yr) is subtracted from annual oil emissions, assuming root/soil carbon sequestration via this biofuel-for-oil substitution is at least as substantial as in Tilman et al. [87]. An additional option that could contribute to this wedge is using biofuels in powerplants with $CO_2$ capture and sequestration [86].

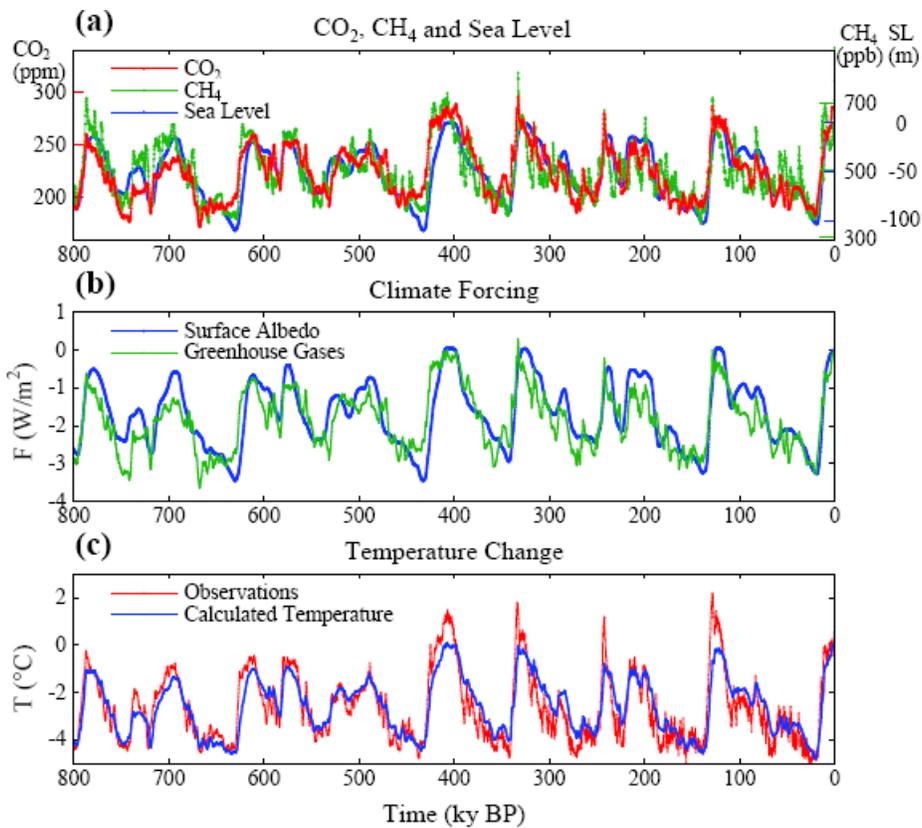

**Fig. (S18). (a)** $CO_2$ [S38], $CH_4$ [S39] and sea level [S16] for past 800 ky. **(b)** Climate forcings due to changes of GHGs and ice sheet area, the latter inferred from the sea level history of Bintanja et al. [S16]. **(c)** Calculated global temperature change based on the above forcings and climate sensitivity ¾°C per $W/m^2$. Observations are Antarctic temperature change from the Dome C ice core [S8] divided by two.

## 16. EPICA 800 ky data

Antarctic Dme C ice core data acquired by EPICA (European Project for Ice Coring in Antarctica) provide a record of atmospheric composition and temperature spanning 800 ky [S8], almost double the time covered by the Vostok data [17, 18] of Figs. **(1)** and **(2)**. This extended record allows us to examine the relationship of climate forcing mechanisms and temperature change over a period that includes a substantial change in the nature of glacial-interglacial climate swings. During the first half of the EPICA record, the period 800-400 ky BP, the climate swings were smaller, sea level did not rise as high as the present level, and the GHGs did not increase to amounts as high as those of recent interglacial periods.

Fig. **(S18)** shows that the temperature change calculated exactly as described for the Vostok data of Fig. **(1)**, i.e., multiplying the fast-feedback climate sensitivity ¾°C per $W/m^2$ by the sum of the GHG and surface albedo forcings (Fig. **S18b**), yields a remarkably close fit in the first half of the Dome C record to one-half of the temperature inferred from the isotopic composition of the ice. In the more recent half of the record slightly larger than ¾°C per $W/m^2$ would yield a noticeably better fit to the observed Dome C temperature divided by two (Fig. **S19**). However, there is no good reason to change our approximate estimate of ¾°C per $W/m^2$, because the assumed polar amplification by a factor of two is only approximate.

The sharper spikes in recent observed interglacial temperature, relative to the calculated temperature, must be in part an artifact of differing temporal resolutions. Temperature is inferred from the isotopic composition of the ice, being a function of the temperature at which the snowflakes formed, and thus inherently has a very high temporal resolution. GHG amounts, in contrast, are smoothed over a few ky by mixing of air in the snow that occurs up until the snow is deep enough for the snow to be compressed into ice. In the central Antarctic, where both Vostok and Dome C are located, bubble closure requires a few thousand years [17].

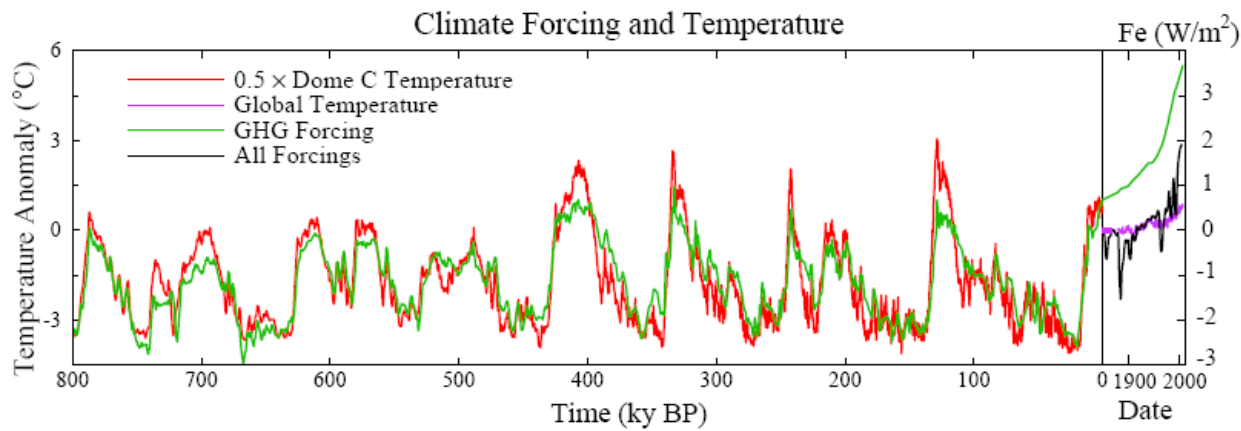

**Fig. (S19).** Global temperature change (left scale) estimated as half of temperature change from Dome C ice core [S8] and GHG forcing (right scale) due to $CO_2$, $CH_4$ and $N_2O$ [S38, S39]. Ratio of temperature and forcing scales is 1.5°C per W/m$^2$. Time scale is expanded in the extension to recent years. Modern forcings include human-made aerosols, volcanic aerosols and solar irradiance [5]. GHG forcing zero point is the mean for 10-8 ky before present. Net climate forcing and modern temperature zero points are at 1850. The implicit presumption that the positive GHG forcing at 1850 is largely offset by negative human-made forcings [7] is supported by the lack of rapid global temperature change in the Holocene [**Fig. (S6)**].

## 17. Comparison of Antarctic data sets

Fig. (**S20**) compares Antarctic data sets used in this supplementary section and in our parent paper. This comparison is also relevant to interpretations of the ice core data in prior papers using the original Vostok data.

The temperature records of Petit et al. [17] and Vimeux et al. [18] are from the same Vostok ice core, but Vimeux et al. [18] have adjusted the temperatures with a procedure designed to correct for climate variations in the water vapor source regions. The isotopic composition of the ice is affected by the climate conditions in the water vapor source region as well as by the temperature in the air above Vostok where the snowflakes formed; thus the adjustment is intended to yield a record that more accurately reflects the air temperature at Vostok. The green temperature curve in Fig. (**S20c**), which includes the adjustment, reduces the amplitude of glacial-interglacial temperature swings from those in the original (red curve) Petit et al. [17] data. Thus it seems likely that there will be some reduction of the amplitude and spikiness of the Dome C temperature record when a similar adjustment is made to the Dome C data set.

The temporal shift of the Dome C temperature data [S8], relative to the Vostok records, is a result of the improved EDC3 [S40, S41] time scale. With this new time scale, which has a 1σ uncertainty of ~3 ky for times earlier than ~130 ky BP, the rapid temperature increases of Termination IV (~335 ky BP) and Termination III (~245 ky BP) are in close agreement with the contention [7] that rapid ice sheet disintegration and global temperature rise should be nearly simultaneous with late spring (April-May-June) insolation maxima at 60N latitude, as was already the case for Terminations II and I, whose timings are not significantly affected by the improved time scale.

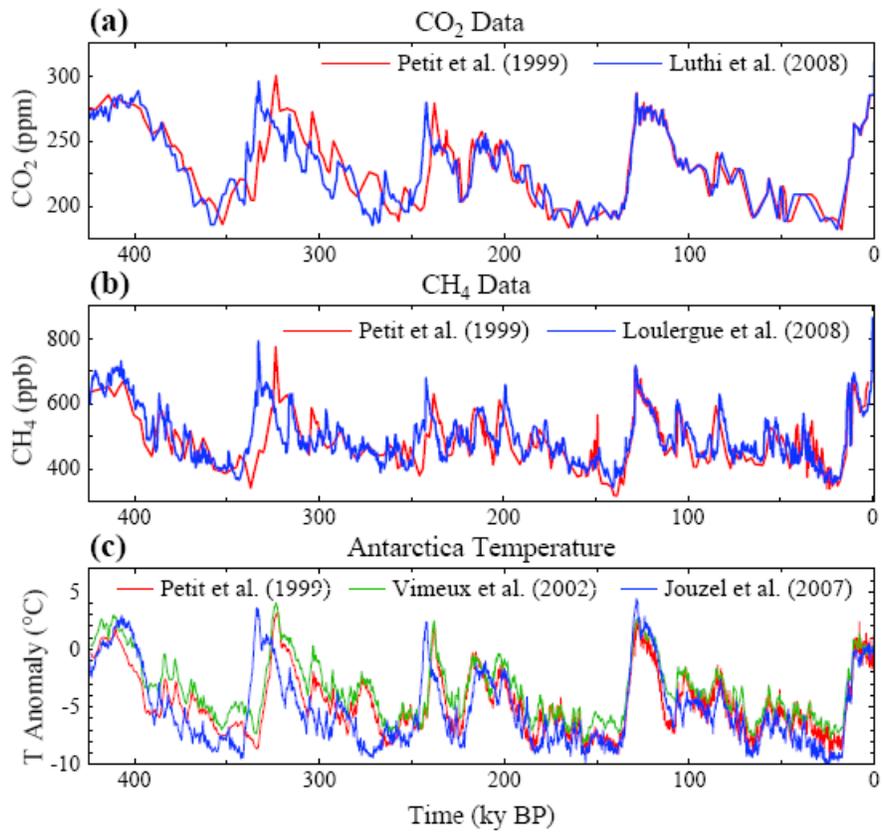

**Fig. (S20).** Comparison of Antarctic $CO_2$, $CH_4$, and temperature records in several analyses of Antarctic ice core data.

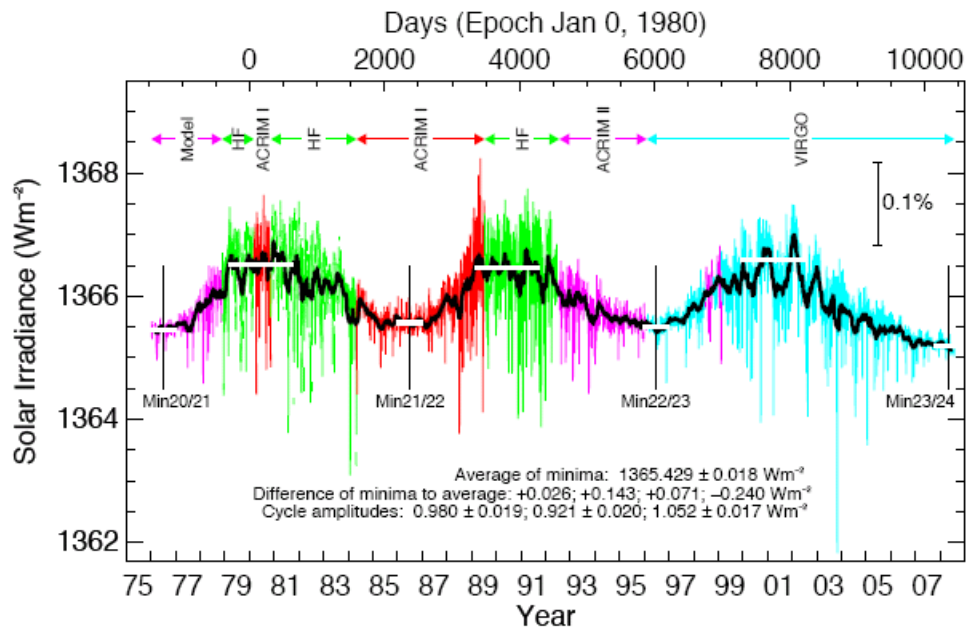

**Fig. (S21).** Solar irradiance from composite of several satellite-measured time series based on Frohlich and Lean [S44].

## 18. Climate variability, climate models, and uncertainties

Climate exhibits great variability, forced and unforced, which increases with increasing time scale [2, 90, 91]. Increasing abilities to understand the nature of this natural variability and improving modeling abilities [S42] do not diminish the complications posed by chaotic variability for interpretation of ongoing global change.

Expectation that global temperature will continue to rise on decadal time scales is based on a combination of climate models and observations that support the inference that the planet has a positive energy imbalance [5, 8, 96]. If the planet is out of energy balance by +0.5-1 W/m$^2$, climate models show that global cooling on decadal time scales is unlikely [96], although one model forecast [95] suggests that the Atlantic overturning circulation could weaken in the next decade, causing a regional cooling that offsets global warming for about a decade.

The critical datum for determining the certainty of continued global warming on decadal time scales is the planet's energy imbalance. Improved evaluations of ocean heat storage in the upper 700 m of the ocean [97] yield ~0.5 x 10$^{22}$ J/yr averaged over the past three decades, which is ~0.3 W/m$^2$ over the full globe. Our model has comparable heat storage in the ocean beneath 700 m, but limited observational analyses for the deep ocean [S43] report negligible heat storage.

If our modeled current planetary energy imbalance of 0.5-1 W/m$^2$ is larger than actual heat storage, the likely explanations are either: (1) the climate model sensitivity of 3°C for doubled $CO_2$ is too high, or (2) the assumed net climate forcing is too large. Our paleoclimate analyses strongly support the modeled climate sensitivity, although a sensitivity as small as 2.5 W/m$^2$ for doubled $CO_2$ could probably be reconciled with the paleoclimate data. The net climate forcing is more uncertain. Our model [8] assumes that recent increase of aerosol direct and indirect (cloud) forcings from developing country emissions are offset by decreases in developed countries.

These uncertainties emphasize the need for more complete and accurate measurements of ocean heat storage, as well as precise global observations of aerosols including their effects on clouds. The first satellite observations of aerosols and clouds with the needed accuracy are planned to begin in 2009 [98]. Until accurate observations of the planetary energy imbalance and global climate forcing are available, and found to be consistent with modeled climate sensitivity, uncertainties in decadal climate projections will remain substantial.

The sun is another source of uncertainty about climate forcings. At present the sun is inactive, at a minimum of the normal ~11 year solar cycle, with a measureable effect on the amount of solar energy received by Earth (Fig. S21). The amplitude of solar cycle variations is about 1 W/m$^2$ at the Earth's distance from the sun, a bit less than 0.1% of the ~1365 W/m$^2$ of energy passing through an area oriented perpendicular to the Earth-sun direction.

Climate forcing due to change from solar minimum to solar maximum is about ¼ W/m$^2$, because the Earth absorbs ~235 W/m$^2$ of solar energy, averaged over the Earth's surface. If equilibrium climate sensitivity is 3°C for doubled $CO_2$ (¾°C per W/m$^2$), the expected equilibrium response to this solar forcing is ~0.2°C. However, because of the ocean's thermal inertia less than half of the equilibrium response would be expected for a cyclic forcing with ~11 year period. Thus the expected global-mean transient response to the solar cycle is less than or approximately 0.1°C.

It is conceivable that the solar variability is somehow amplified, e.g., the large solar variability at ultraviolet wavelengths can affect ozone. Indeed, empirical data on ozone change with the solar cycle and climate model studies indicate that induced ozone changes amplify the direct solar forcing, but amplification of the solar effect is by one-third or less [S44, S45].

Other mechanisms amplifying the solar forcing have been hypothesized, such as induced changes of atmospheric condensation nuclei and thus changes of cloud cover. However, if such mechanisms were effective, then an 11-year signal should appear in temperature observations (Fig. 7). In fact a very weak solar signal in global temperature has been found by many investigators, but only of the magnitude (~0.1°C or less) expected due to the direct solar forcing.

The possibility remains of solar variability on longer time scales. If the sun were to remain 'stuck' at the present solar minimum (Fig. S21) it would be a decrease from the mean irradiance of recent decades by ~0.1%, thus a climate forcing of about -0.2 W/m$^2$.

The current rate of atmospheric $CO_2$ increase is ~2 ppm/year, thus an annual increase of climate forcing of about +0.03 W/m$^2$ per year. Therefore, if solar irradiance stays at its recent minimum value, the climate forcing would be offset by just seven years of $CO_2$ increase. Human-made GHG climate forcing is now increasing at a rate that overwhelms variability of natural climate forcings.

Climate models are another source of uncertainty in climate projections. Our present paper and our estimated target $CO_2$ level do not rely on climate models, but rather are based on empirical evidence from past and ongoing climate change. However, the limited capability of models to simulate climate dynamics and interactions among climate system components makes it difficult to estimate the speed at which climate effects will occur and the degree to which human-induced effects will be masked by natural climate variability.

The recent rapid decline of Arctic ice [S47-S49] is a case in point, as it has been shown that model improvements of multiple physical processes will be needed for reliable simulation. The modeling task is made all the more difficult by likely connections of Arctic change with the stratosphere [S50] and with the global atmosphere and ocean [S51].

## APPENDIX REFERNCES